\documentclass[conference,a4paper]{IEEEtran}
\IEEEoverridecommandlockouts

\newcommand{\mywidetilde}[1]{\widetilde{#1}}
\newcommand{\mywidebar}[1]{\bar{#1}}
\newcommand{\mywidehat}[1]{\hat{#1}}

\renewcommand{\mywidetilde}[1]{\accentset{\sim}{#1}}

\usepackage{balance}
\usepackage{moresize}
\usepackage{booktabs}
\usepackage{bm,amsmath,amssymb,amsfonts}
\usepackage{graphicx}
\usepackage{textcomp}
\usepackage{xcolor}
\def\BibTeX{{\rm B\kern-.05em{\sc i\kern-.025em b}\kern-.08em
    T\kern-.1667em\lower.7ex\hbox{E}\kern-.125emX}}
    
\usepackage{academicons}
\definecolor{orcidlogocol}{HTML}{A6CE39}

\usepackage[backend=biber,style=ieee, citestyle=numeric, sortcites=true,isbn=false, doi=false, url=false,mincitenames=1, maxcitenames=1,maxbibnames=3]{biblatex}
\usepackage[detect-all=true, binary-units=true]{siunitx}
\usepackage{array}
\usepackage{booktabs}
\usepackage[hidelinks]{hyperref}
\usepackage{algorithm,algorithmic}

\usepackage{multicol,multirow}

\usepackage{accents}

\usepackage{soul}
\soulregister\cite7
\soulregister\ref7
\soulregister\SI7
\soulregister\ac7

\usepackage[detect-all=true, binary-units=true]{siunitx}
\DeclareSIUnit\decibelm{dBm}

\usepackage[nolist]{acronym}
\acused{wifi}
\acused{wlan}
\acused{gnss}

\newcommand{\MYheader}{2021 International Conference on Indoor Positioning and Indoor Navigation (IPIN), 29
Nov. -- 2 Dec. 2021, Lloret de Mar, Spain\\}

\newcommand{\mycp}{978-1-6654-0402-0/21/\$31.00~\copyright~2021~IEEE\hfill}

\renewcommand{\MYheader}{}
\renewcommand{\mycp}{}

\makeatletter
\def\ps@headings{%
	\def\@oddhead{\MYheader}
	\def\@evenhead{\MYheader}
	\def\@oddfoot{}%
	\def\@evenfoot{}}
\def\ps@IEEEtitlepagestyle{%
	\def\@oddhead{\MYheader}
	\def\@evenhead{\MYheader}
	\def\@oddfoot{}%
	\def\@evenfoot{\mycp}
    }
\makeatother
\begin{document}

\title{Towards Ubiquitous Indoor Positioning:\\ Comparing Systems across Heterogeneous Datasets%
\thanks{%
Corresponding Author: J. Torres-Sospedra (\texttt{torres@ubikgs.com})}
\thanks{The authors gratefully acknowledge funding from European Union’s Horizon 2020 Research and Innovation programme under the Marie Sklodowska Curie grant agreement No. $813278$ (A-WEAR: A network for dynamic wearable applications with privacy constraints, http://www.a-wear.eu/). 
FCT -- Fundação para a Ciência e Tecnologia within the R\&D Units Project Scope: UIDB/00319/2020 and the PhD fellowship PD/BD/137401/2018. J. Torres-Sospedra acknowledges funding from MICIU (INSIGNIA, PTQ2018-009981)}}

\author{\IEEEauthorblockN{
 Joaquín Torres-Sospedra\IEEEauthorrefmark{1}, %
 Ivo Silva\IEEEauthorrefmark{2}, %
 Lucie Klus\IEEEauthorrefmark{3}\IEEEauthorrefmark{4},
 Darwin Quezada-Gaibor\IEEEauthorrefmark{3}\IEEEauthorrefmark{4}, %
 Antonino Crivello\IEEEauthorrefmark{5},\\
 Paolo Barsocchi\IEEEauthorrefmark{5}, 
 Cristiano Pendão\IEEEauthorrefmark{2},
 Elena Simona Lohan\IEEEauthorrefmark{3}, %
 Jari Nurmi\IEEEauthorrefmark{3}, %
 and Adriano Moreira\IEEEauthorrefmark{4} %
}
\IEEEauthorblockA{%
\IEEEauthorrefmark{1}\textit{UBIK Geospatial Solutions S.L.}, Castellon, Spain\\
\IEEEauthorrefmark{2} \textit{Algoritmi Research Center}, \textit{University of Minho}, Guimarães, Portugal\\
\IEEEauthorrefmark{3}\textit{Faculty of Information Technology and Communication Sciences}, \textit{Tampere University}, Tampere, Finland\\
\IEEEauthorrefmark{4}\textit{Institute of New Imaging Technologies}, \textit{Universitat Jaume I}, Castellon, Spain\\
\IEEEauthorrefmark{5}\textit{Information Science and Technologies Institute}, \textit{National Research Council}, Pisa, Italy
}}

\maketitle

\begin{abstract}The evaluation of \acfp{ips} mostly relies on local deployments in the researchers' or partners' facilities. The complexity of preparing comprehensive experiments, collecting data, and considering multiple scenarios usually limits the evaluation area and, therefore, the assessment of the proposed systems.
The requirements and features of controlled experiments cannot be generalized since the use of the same sensors or anchors density cannot be guaranteed. The dawn of datasets is pushing \ac{ips} evaluation to a similar level as machine-learning models, where new proposals are evaluated over many heterogeneous datasets. This paper proposes a way to evaluate \acp{ips} in multiple scenarios, that is validated with three use cases. The results prove that the proposed aggregation of the evaluation metric values is a useful tool for high-level comparison of \acp{ips}. \end{abstract}

\begin{IEEEkeywords}
Evaluation; Indoor Positioning Benchmarking
\end{IEEEkeywords}

\begin{acronym}[XXX] 
\acro{ap}[AP]{Access Point}
\acro{akm}[AkM]{Adaptive k-Means}
\acro{ble}[BLE]{Bluetooth Low Energy}
\acro{cr}[CR]{Compression Ratio}
\acro{fp}[FP]{fingerprinting}
\acro{lbs}[LBS]{location-based service}
\acro{gnss}[GNSS]{Global Navigation Satellite System}
\acro{iot}[IoT]{Internet of Things}
\acro{ips}[IPS]{Indoor Positioning System}
\acro{mac}[MAC]{Media Access Control}
\acro{ml}[ML]{Machine Learning}
\acro{mse}[MSE]{Mean Square Error}
\acro{nn}[NN]{Nearest Neighbour}
\acro{pmlb}[PMLB]{Penn Machine Learning Benchmark}
\acro{rf}[RF]{Radio Frequency}
\acro{rp}[RP]{Reference Point}
\acro{rs}[RS]{Recommender Systems}
\acro{rss}[RSS]{Received Signal Strength}
\acro{uci}[UCI]{University of California, Irvine}
\acro{weka}[WEKA]{Waikato Environment for Knowledge Analysis}
\acro{wifi}[\mbox{Wi-Fi}]{IEEE 802.11 Wireless LAN}
\acro{wlan}[WLAN]{Wireless LAN}
\acro{wsn}[WSN]{Wireless Sensors Networks}
\acro{ipin}[IPIN]{Indoor Positioning and Indoor Navigation}
\end{acronym}

\section{Introduction}
\label{sec:introduction}

In the last decade, \acp{ips} have attracted interest from researchers and industries showing, nowadays, good performance and accuracy in different scenarios and test cases. In literature, especially reading papers from specific conferences and scientific journals, the evolution of these systems is quite clear. Starting from the first works in this field, researchers have shown several techniques and technologies able to accurately estimate a target position, e.g., people or end-users devices. For example, by examining the proceedings of the \ac{ipin} conference, a reader can clearly observe that, up to now, several efforts have been made to create common evaluation frameworks and to set common scenarios to increase the chance of generalizing the results obtained by researchers. With these considerations in mind, it is worth noting the valuable results of the \ac{ipin} Competitions \cite{renaudin2019evaluating,potorti2020ipin} in which organizers set several tracks in the same scenarios, offering a common testbed to competitors in order to evaluate their own systems. 

Organizers of the IPIN Competitions have also proposed an evaluation framework \cite{potorti2017comparing}, nowadays widely adopted by the research community for evaluating online and offline systems. With the same goal of offering common data and scenarios for evaluation purposes, several researchers have proposed in the last years free and accessible datasets collected in different environments (e.g. hospitals~\cite{10.1371/journal.pone.0205392}, universities~\cite{german_dataset}, malls~\cite{lopez_dataset}, or factories~\cite{moreira_dataset}) and exploiting different positioning technologies. 

All those efforts have led to systems and technologies quite stable to hit the market. Nevertheless, the main challenge in this field is to be able to generalize the results and the methods of the systems in heterogeneous environments. In other words, systems are starting to show robust performances when deployed in a specific scenario, but their performances may drop significantly if deployed in a different scenario. In literature, \acp{ips} are shown and tested in real-world scenarios but they are generally developed and tuned to obtain the best performance (e.g., the positioning accuracy) in the considered testing environments. This customization reduces the generalization ability of an IPS to work in any scenario. 

The huge availability of public datasets could help researchers to find the best system settings, but typically the \acp{ips} are still only evaluated in one or two scenarios (see Fig.~\ref{fig:relatedwork}). Furthermore, we remark that a shared framework for evaluating over multiple datasets could improve the evaluation process. For example, such a framework can increase the trustiness in sharing scientific results and can enable research reproducibility. We also remark that, a consensus on which features are important during the evaluation process is still missing. In fact, if the overall accuracy is an important metric, the execution time and the computational cost are also fundamental variables for \acp{ips} which have the ambition to overcome the research grade. The main contributions of this paper can be summarized as follows:  
\begin{itemize}
    \item We introduce a novel way to aggregate the evaluation metrics, considering different, heterogeneous, scenarios;
    \item We propose guidelines and recommendations for guiding researchers in evaluating their positioning systems;
    \item Through use cases we validate our proposal showing how multiple datasets can improve the \acp{ips} evaluation.
\end{itemize}




\section{Related work}
\label{sec:relatedwork}

Traditionally, the evaluation of novel \acf{ml} models has included a large setup with multiple diverse datasets, i.e., the evaluation is not limited to just one problem, and it incorporates several datasets covering a heterogeneous set of problems including, for instance, the identification of iris plants, prediction of whether an income exceeds an amount based on census data, the origin of wines, or detection of the presence of a heart disease in a patient, among many others. The traditional datasets can be found in the \ac{uci} machine-learning repository~\cite{uci}. 

Bradley investigated the use of the  area  under  the  receiver  operating  characteristic (ROC) curve  (AUC)  as  a  performance  measure for \ac{ml} algorithms in \cite{BRADLEY19971145}. The metric evaluation included a comparison with six \ac{ml} models and six datasets from \ac{uci}. Datasets were diverse and covered issues on post-operative bleeding, breast cancer, diabetes and heart disease. 

Yang et al.~\cite{982883} introduced a survey of face-recognition models where they identified nine datasets valid for training purposes (where each photo contained just one individual) and four datasets for testing purposes (presenting several challenges for face recognition). The datasets were independently collected by Kodak, Harvard University, Yale University, AT\&T, and MIT among others. The authors of the survey identified several weaknesses, namely, evaluation using a modest-sized standard test set, “tweaking” the models to get better performance on the test set or even testing on the training set, which is an unacceptable practice in \ac{ml}. Some of these weaknesses have been often detected in the evaluation of \acp{ips}. For instance, collecting consecutive \ac{rss} fingerprints and directly splitting the collected dataset into training and testing with cross-validation might be considered data leaking. The resulting test set would not be fully independent, driving to over-optimistic accuracy. If the operating system buffering is not taken into account, the same fingerprint vector may end up in the training and test sets.
Yang et al. concluded that the fair and effective performance evaluation requires careful design of protocols, scope, and, above all, datasets.

Despite the fact that \ac{ml} models are usually evaluated with a ``moderate'' number of datasets as in~\cite{LIN20081817}, it is not unusual to find works where the evaluation considers a very large set of diverse datasets. Fernández-Delgado et al.~\cite{JMLR:v15:delgado14a} proposed an evaluation with $121$ datasets which was later adopted by Zhang et al.~\cite{ZHANG20161094}. Hynes et al.~\cite{Hynes2017TheDL} provided an evaluation over $600$ databases hosted on Kaggle. 
%
%
%
%
However, a review on \ac{ml}~\cite{9000651} has shown that most of recent works are evaluated with one or two datasets, and that an  evaluation with three or more datasets is less frequently encountered in the current literature. The new trends on deep learning applied to particular problems and the existence of extremely large datasets (with millions of samples) have driven the ML algorithms to computationally demanding evaluation procedures. However, in localization domain, an evaluation considering many ``traditional'' moderate-sized datasets is still possible~\cite{9391692}.


Olson et al.~\cite{Olson2017} provided a comparison of a few \ac{ml} models using $165$ real-world curated datasets from the \ac{pmlb} suite, which currently has has $299$ datasets (April 2021). 
Several useful ways to summarize the full results as images were also provided. Other \ac{ml} tools --such as WEKA
, \emph{scikit-learn} 
, TensorFlow 
 or Keras%
-- are easing the integration of \ac{ml} models in real-world implementations.

However, the level of evaluation carried out in the \ac{ml} domain is often not reached in the indoor-positioning area, which usually relies on the evaluation of a single setup with very controlled conditions. Collecting data for wireless indoor positioning is a time-consuming and demanding procedure. However, there are many attempts to provide public datasets in this domain. Fruit of those datasets, Saccomanno et al.~\cite{9391692} provided a comprehensive study where the relation between Wi-Fi fingerprints and the spatial knowledge was explored for indoor positioning using multiple datasets, which extended the setup previously provided by Torres-Sospedra et al. in~\cite{9115419,9169843}. 

\begin{figure}[!h]
    \centering
    \includegraphics[width=0.9\columnwidth,trim={0.303cm 0.833cm  0.3580cm 0.4270cm},clip]{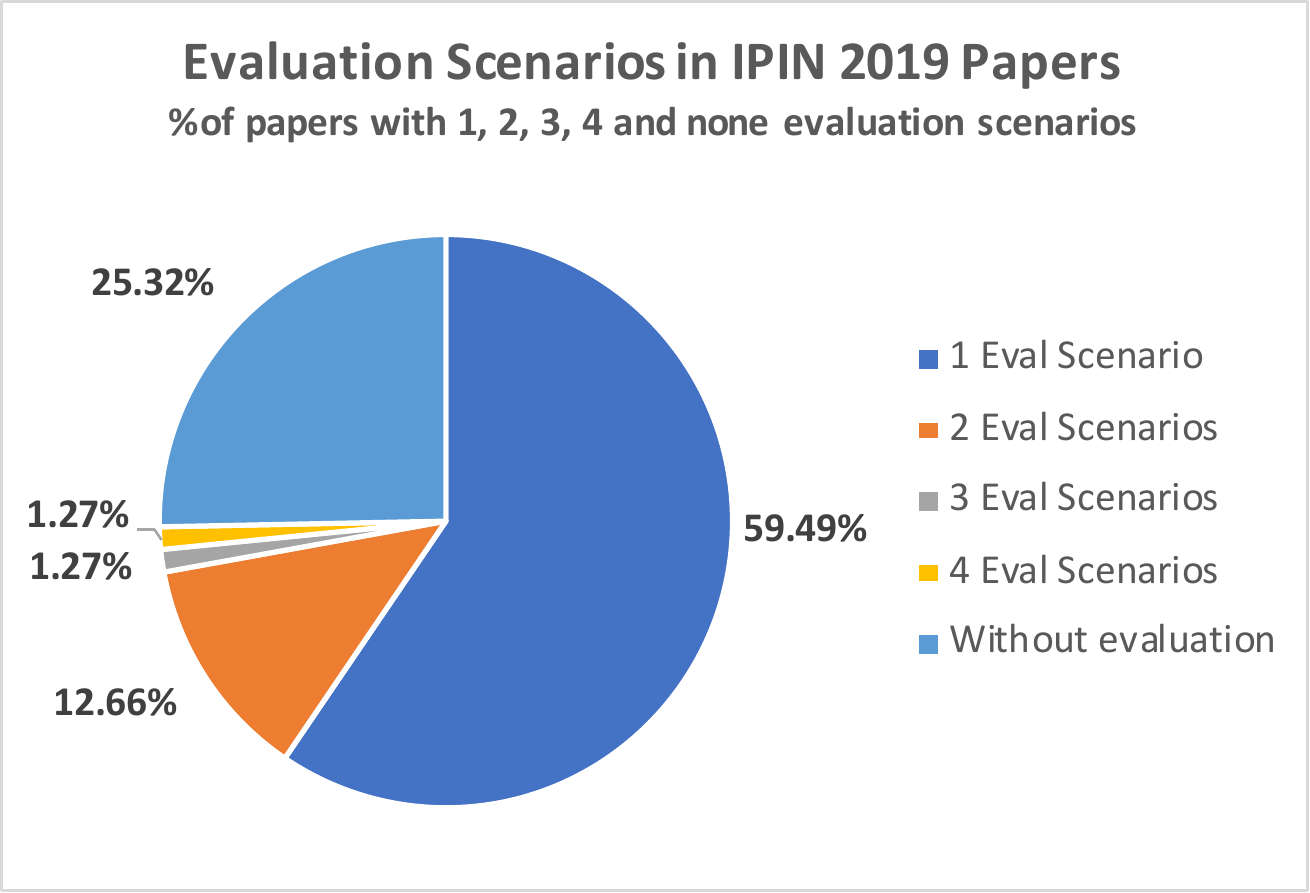}
    \caption{Analysis of the evaluation of the regular papers presented in the IPIN 2019 Conference}
    \label{fig:relatedwork}
\end{figure}

Although there are several datasets and database repositories~\cite{8115940,8911748} available for \ac{ips} evaluation, most research papers still use their own closed setups or datasets. The $79$ accepted papers of the IPIN 2019 conference were analysed to have a better picture of the current trends in evaluating \ac{ips}. The results of the review are shown in Fig.~\ref{fig:relatedwork}, where it can be seen that most of the works with empirical evaluation only included 1 or 2 scenarios, and only two papers included 3~\cite{8911762} and 4~\cite{8911799} scenarios respectively. These data lead to remark the importance of finding a common benchmark for evaluation purpose. The efforts in terms of standardization are increasing and, in this context, guidelines and strategies for comparing \ac{ips} in several and heterogeneous scenarios represent a step forward in this research field. In fact, in order to understand which technologies and techniques are more able to fit in different environments, we should promote the adoption of common datasets, and as future works we should probably try to standardizing them, to promote reliable and robust systems.










In this paper, we would like to mimic the evaluation procedures in \ac{ml}, this time applied to \acp{ips}, and  show the importance of evaluating IPSs with multiple diverse datasets with the proposed aggregated metrics and visualization tools.

\clearpage
\section{Materials and Methods}
\label{sec:methoddescription}
This section focuses on the proposed methodology to aggregate metrics and to how perform visual comparisons.
\subsection{Aggregating evaluation metrics}

Let's suppose that we follow the Black-box testing approach suggested in the ISO18305 \cite{iso18305} and discussed in \cite{potorti2018evaluation}. Given an IPS, its evaluation metric can be represented by:
\begin{equation}
 \mathcal{M}^{scenario,trial}_{method}
\end{equation}
where $\mathcal{M}$ represents the evaluation metric (e.g., mean positioning error, third quartile error, root mean squared error, floor hit rate, execution time, among many others), $method$ identifies the evaluated method, $scenario$ corresponds to the evaluation scenario and $trial$ corresponds to the trial number (or execution run) for that  scenario. The best value for a metric depends on its nature. For instance, developers want to provide \ac{ips} with low positioning error and high floor detection rate.

It is worth noting that in a dataset-based evaluation, the scenario corresponds to the dataset itself, whereas the trial corresponds to the execution run. In an on-line evaluation, without datasets, the scenario can be considered the combination of the evaluation area and the positioning infrastructure. 

In the simplest evaluation, with a single metric, scenario and trial, several methods can be directly compared, i.e. the method reporting the best metric value can be considered the best solution. However, real-world evaluation may include multiple runs or trials under the same scenario as the \ac{ips} might be affected by environmental conditions or a random initialization. For instance, in the \ac{ipin} annual competition, the participants are able to provide multiple runs being the trial with providing the best accuracy the used for ranking. In other domains, such as machine learning where some models depend on random initialization, the average over the multiple runs is applied to obtain the final value for the metric.
In this paper, we aggregate the results for multiple trials as the average among all the $N_{trials}$ number of trials.
\begin{equation}
 \mywidebar{\mathcal{M}}^{scenario}_{method} =
 \frac{\sum^{N_{trials}}_{t = 1}{\mathcal{M}^{scenario,t}_{method}}}{N_{trials}}
\end{equation}
As the metric values usually depend on the scenario, we propose to normalize the metric to a base line becoming the unitless metric. That normalization should be done with respect to a simple method or configuration. In \ac{rss}-based fingerprinting models, the baseline could be the $1$-NN algorithm.
\begin{equation}
 \mywidehat{\mathcal{M}}^{scenario}_{method} =
 \frac{\mywidebar{\mathcal{M}}^{scenario}_{method}}
 {\mathcal{M}^{scenario}_{baseline}}
\end{equation}
In addition, a comprehensive evaluation should include different scenarios covering multiple cases, since an indoor positioning solution may behave differently in two different scenarios. To integrate different scenarios, we propose to report the aggregated-values average and the standard deviation of the baseline-normalized values for all scenarios as follows:
\begin{equation}
 \mywidetilde
 {\mathcal{M}}_{method} =
 \frac{\sum_{scenario = 1}^{N_{scenarios}}{\mywidehat{\mathcal{M}}^{scenario}_{method}}}
 {N_{scenarios}}
\end{equation}
The average of the baseline-normalized values is providing the general behaviour of the method considering multiple scenarios, whereas the standard deviation reflects the variability of the metric along all the scenarios considered. When evaluating an \ac{ips}, we target those methods providing the best averaged value with the lowest possible deviation. It is worth noting that here we use the term ``best'' on purpose as there are metrics, such as the ones based on the positioning error, where the averaged values should be as lowest as possible. On the other hand in metrics such as the floor identification rate (percentage of correctly identified floor number), the averaged values must be as high as possible. 

As the on-line evaluation of \acp{ips} is very demanding, we will integrate the proposed approach to aggregate the results off-line. Thus, the \acp{ips} will be evaluated with pre-recorded datasets with independent dedicated training and evaluation sets. Using the same dataset over the different trials should not affect the evaluation metrics based on the positioning error, as the data used is the same in the multiple runs. However, if the method employed relied on a kind of random initialization (such as a Neural Network), then the positioning error should vary from run to run.  

\subsection{Comparing two different metrics over multiple datasets}

Evaluation becomes more complex when several targets must be accomplished. For instance, one could aim simultaneously at providing the lowest possible 3D positioning error and the lowest execution time. With the proposed way to aggregate a metric over different scenarios and trials, the simplest option is to provide the information as a table with as many columns as metrics considered. If the number of metrics is two, it can be complemented with a scatter plot showing the average of the baseline-normalized values for the two metrics.  

\begin{figure}[!hb]
        \centering
    \scriptsize
    \tabcolsep 6pt
    \begin{tabular}{ccccc}
\includegraphics[width=0.105\columnwidth]{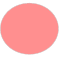} & \includegraphics[width=0.105\columnwidth]{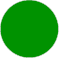}& \includegraphics[width=0.105\columnwidth]{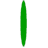}& \includegraphics[width=0.105\columnwidth]{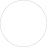}\\
$\mywidetilde{\epsilon_{\text{3D}}} = 0.88$&
$\mywidetilde{\epsilon_{\text{3D}}} = 0.99$&
$\mywidetilde{\epsilon_{\text{3D}}} = 7.18$&
$\mywidetilde{\epsilon_{\text{3D}}} = 1.00$\\
$\mywidetilde{\tau_{\text{DB}}} = 1.45$&
$\mywidetilde{\tau_{\text{DB}}} = 0.09$&
$\mywidetilde{\tau_{\text{DB}}} = 0.16$&
$\mywidetilde{\tau_{\text{DB}}} = 1.00$\\

    \end{tabular}
    \caption{Example of elements in the GMMS plot}
    \label{fig:resultsexample}
\end{figure}

If one would like to go a step further and show the results for each dataset with the two metrics, we propose a novel graphical representation (we call it a GMMS plot), which provides four dimensions in a single plot. The x-axis corresponds to the scenario (or dataset), the y-axis to the method and each plotted element is a colored ellipse whose color (green range, white and red range) indicates one evaluation metric and the shape (horizontal, circled, vertical) stands for the other metric. Fig.~\ref{fig:resultsexample} shows an example on how elements are displayed in a GMMS plot according to the aggregated mean values of positioning error $\mywidetilde{\epsilon_{\text{3D}}}$ and dataset execution time, $\mywidetilde{\tau_{\text{DB}}}$. With respect to the baseline, the more red the element is the worse (higher value) $\mywidetilde{\tau_{\text{DB}}}$ is, in the same way the more vertical the element is the higher $\mywidetilde{\epsilon_{\text{3D}}}$ is.
On the contrary, the greener the element is, the lower  $\mywidetilde{\tau_{\text{DB}}}$ is and, in the same way, the more horizontal the element is, the lower $\mywidetilde{\epsilon_{\text{3D}}}$ is. For the baseline values, we use a white circle.

\section{Use cases}
\label{sec:usecases}

\subsection{Analysis on the parameters of the $k$-NN algorithm}

As the first use case, we provide a general analysis of the distance metrics for the $k$-\ac{nn} algorithm used in fingerprinting. The intention is not to analyse the algorithm, but to show the potential of the proposed aggregation metrics to perform a more general comparison. We have considered 16 public as in~\cite{9115419,9169843,9391692} and the experiments were run in a computer with Intel  Core  i7-8700 CPU and  Octave 4.0.3. Moreover, we consider two evaluation metrics the mean positioning error, $\epsilon_{\text{3D}}$, and the dataset execution time, $\tau_{\text{DB}}$. Due to the lack of space, we only show the aggregated metrics and comment on particular results. 


First, the analysis was run using the $k$-\ac{nn} algorithm with the following configuration: $k = \num{1}$, positive data representation~\cite{torressospedra2015comprehensive} and the city block distance as distance/similarity metric. This plain version of the $k$-\ac{nn} can be considered our baseline for further comparisons, whose full results are reported in Table~\ref{table:cost1nn}. Please note that the metrics $\mywidebar{\epsilon_{\text{3D}}}$ and $\mywidebar{\tau_{\text{DB}}}$ provides the averaged values over 10 runs, whereas the metrics $\mywidehat{\epsilon_{\text{3D}}}$ and $\mywidehat{\tau_{\text{DB}}}$ are normalized to the baseline. The aggregated metrics $\mywidetilde{\epsilon_{\text{3D}}}$ and $\mywidetilde{\tau_{\text{DB}}}$ provides the average and standard dev. of the normalized values across the 16 datasets. 

\begin{table}[hbtp]
\centering
\caption{Full results of 1-NN, city block distance and positive data rep.}
\label{table:cost1nn}
\begin{tabular}{lSSccc}   
\toprule
&\multicolumn{2}{c}{Absolute values}&
&\multicolumn{2}{c}{Norm. values}\\\cmidrule{2-3}\cmidrule{5-6}
Dataset& 
{$\mywidebar{\epsilon_{\text{3D}}}$ (m)} & {$\mywidebar{\tau_{\text{\text{DB}}}}$ (s)} && {$\mywidehat{\epsilon_{\text{3D}}}$} & {$\mywidehat{\tau_{\text{DB}}}$}  
\\\midrule
DSI~1&    4.95&   12.21   &&1&1\\
DSI~2&    4.95&    5.15   &&1&1\\
LIB~1&    3.02&   46.19   &&1&1\\
LIB~2&    4.18&   46.39   &&1&1\\
MAN~1&    2.82&  155.46   &&1&1\\
MAN~2&    2.47&   14.26   &&1&1\\
SIM  &    3.24&  252.00   &&1&1\\
TUT~1&    9.59&   18.93   &&1&1\\
TUT~2&   14.37&    2.73   &&1&1\\
TUT~3&    9.59&   79.73   &&1&1\\
TUT~4&    6.36&   79.88   &&1&1\\
TUT~5&    6.92&   11.88   &&1&1\\
TUT~6&    1.94&  620.72   &&1&1\\
TUT~7&    2.69&  511.70   &&1&1\\
UJI~1&   10.81&  599.04   &&1&1\\
UJI~2&    8.05& 2924.69   &&1&1\\
\toprule
&&&&
{$\mywidetilde{\epsilon_{\text{3D}}}$} mean(std) & {$\mywidetilde{\tau_{\text{DB}}}$ mean(std)}\\
\midrule
\multicolumn{4}{r}{Plain $k$-NN (baseline)}  & 1.00 (0.00) & 1.00 (0.00)\\
\bottomrule
\end{tabular}
\end{table}

On the one hand, it can be clearly observed that the execution time of the entire evaluation dataset ($\tau_{\text{DB}}$) highly depends on the dataset, as $k$-\ac{nn} computational cost depends on the number of training and evaluation samples. On the other hand, the mean positioning error varies, ranging from almost \SI{2}{\metre} (TUT~6), to more than \SI{14}{\metre} (TUT~2). This variability on the timing and accuracy measurements might make a direct comparison difficult. For example, a reduction of \SI{50}{\centi\metre} in the positioning error is more significant in dataset TUT~6 than in TUT~2. Similarly, a reduction of \SI{2}{\second} in the execution time is more significant in dataset TUT~2 than in TUT~6.

Second, the analysis on the distance function used to compare two fingerprints is shown in Table~\ref{table:cost1nn-distances}. We provide the aggregated positioning error $\tilde{\epsilon_{3D}}$ and the execution time $\tilde{\tau_{DB}}$ of all the alternatives. For both metrics, we provide the average and the standard deviation of the baseline-normalized values over the 16 datasets. The distance metrics were evaluated keeping the other two parameters of the baseline configuration unaltered (positive data representation and $k = 1$). 

\begin{table}[htbp]
\centering 
\caption{Comparison of the aggregated values for the positioning error and execution time for $1$-nn with 14 distance metrics.}
\label{table:cost1nn-distances}
\tabcolsep 0.5pt
\begin{tabular}{l@{\hspace{5pt}} r@{.}l r@{.}l @{\hspace{5pt}} r@{.}l r@{.}l l@{\hspace{10pt}} l@{\hspace{5pt}} r@{.}l r@{.}l @{\hspace{5pt}} r@{.}l r@{.}l}   
\toprule
Distance&\multicolumn{4}{c}{$\mywidetilde{\epsilon_{\text{3D}}}$}& \multicolumn{4}{c}{$\mywidetilde{\tau_{\text{DB}}}$}&
&Distance&\multicolumn{4}{c}{$\mywidetilde{\epsilon_{\text{3D}}}$} & \multicolumn{4}{c}{$\mywidetilde{\tau{_\text{DB}}}$}\\
\cmidrule{1-9}\cmidrule{11-19}

Kulczynski\textsubscript{d}&    0&90&(0&12)&    1&84&(0&02)&& City Block&    1&00&(0&00)&    1&00&(0&00)\\
Kulczynski\textsubscript{s}&    0&90&(0&12)&    1&86&(0&01)&& LGD&           1&00&(0&22)&    2&09&(0&25)\\
Motyka&                         0&90&(0&12)&    1&18&(0&01)&& PLGD10&        0&91&(0&15)&    3&15&(0&32)\\
Ruzicka&                        0&90&(0&12)&    1&30&(0&01)&& PLGD40&        0&95&(0&19)&    3&15&(0&32)\\
Soergel&                        0&90&(0&12)&    1&29&(0&01)&& Euclidean&     0&99&(0&05)&    1&02&(0&01)\\
S\o{}rensen&                       0&90&(0&12)&    1&16&(0&01)&& Neyman&        1&28&(0&36)&    1&59&(0&04)\\
Tanimoto&                       0&90&(0&12)&    1&48&(0&01)&& Euclidean\textsuperscript{2}&    0&99&(0&05)&    0&92&(0&01)\\
\bottomrule
\end{tabular}
\end{table}

The left side of Table~\ref{table:cost1nn-distances} shows the results on some distances that are equivalent between them in terms of sorting the reference samples by distance to the operational sample, and therefore they provide the same positioning errors (\SI{10}{\%} lower than in the baseline for all of them). However, they differ in terms of computational costs, with an increase in costs ranging from \SIrange{16}{86}{\%} on average. Among these equivalent metrics, the S\o{}rensen distance is the one reporting the best computational costs (only \SI{16}{\%} higher with respect to the baseline). In general, the S\o{}rensen distance is reporting lower positioning errors than the Euclidean distance (baseline distance metrics) in most of the datasets, providing a mean positioning error \SI{25}{\%} lower than the baseline for TUT~1.

The right side of Table~\ref{table:cost1nn-distances} shows the results on the remaining metrics. Although the Euclidean distance and city block are not equivalent, they are providing similar general results in terms of averaged normalized positioning error when considering all the datasets (0.99 and 1.0 respectively) and computational costs (1.02 and 1.0 respectively). Despite the former is performing one square per \ac{ap} and one square root operations and the later computes one absolute value per \ac{ap}, their computational times are almost the same in all datasets ($\tau_{\text{DB}} = 1 \pm 0.05$). Despite these similarities in the averaged case, their performance clearly depends on the dataset (for UJI~1 the Euclidean distance is providing an error \SI{12}{\%} lower than the city block distance, but it is \SI{7}{\%} higher for TUT~6). The Squared Euclidean (Euclidean\textsuperscript{2} in Table~\ref{table:cost1nn-distances}) is equivalent to the Euclidean distance in terms of ranking samples by distance but it has a lower computational cost ($1.02$ vs $0.92$), since the square root operation is not needed to obtain an equivalent samples ranking. The three Log Gaussian-based distances (LGD, PLGD10 and PLGD40) have attached a significant increase of the computation time, but in some cases they provide a great improvement on the positioning error (e.g., TUT~1 and PLGD40, where the error has been reduced a \SI{45}{\%} with respect to the baseline). 

In the election of the best distance metric, some concerns about the computational costs might raise. For instance, PLDG40 is better than S\o{}rensen for dataset SIM, their difference in the normalized positioning error is just \SI{2}{\%} (around \SI{7}{cm} if we consider the absolute positioning errors) but the normalized computational costs are very different, $3.15$ and $1.16$ respectively. This means that to reach similar averaged accuracy, the process to estimate the position takes more than two times with PLGD40 than with S\o{}rensen distance and the gain in accuracy might be considered marginal. In our opinion, the election of the best alternative should balance both metrics. In case of similar positioning error, we should select the one that is computationally efficient (green computing). 

Finally, this analysis has shown that the distance/similarity function does not only impact the positioning accuracy but also the computational burden. The average of the baseline-normalized values for all datasets provides the general behavior. The standard deviation identifies where there exists a huge dependency on the dataset and dataset-based analysis is needed to select the optimal value for the parameter. 
\subsection{Comparison on clustering models for Wi-Fi fingerprinting}

The second use case corresponds to the comparison of clustering models to make faster the estimation of the indoor position using $k$-NN algorithm. It is well known that $k$-NN does not require a training phase but, in contrast, it is inefficient as it needs to compute the distance/similarity function between the operational fingerprint and all the reference fingerprints in the radio map. 

One alternative to alleviate the computational burden is to apply clustering models to the radio map, by generating clusters that group fingerprints with similar features. Then, in the operational phase one has to first search for the most similar group (cluster) and then compute the distance function to all the reference fingerprints falling into that cluster. 

For the experiments, we have used the 16 datasets from the first use case and the experiments were carried out on the same desktop computer. The configuration of the $k$-NN estimator for all the methods corresponds to the baseline previously used with $k$ equal to $\num{1}$, the positive data representation and the city block distance as similarity measure for fingerprint comparison. We consider two evaluation metrics: the mean positioning error, $\epsilon_{3D}$ and the dataset execution time, $\tau_{DB}$ to assess the performance of the clustering models.

\begin{figure}[!htb]
    \centering
    \includegraphics[width=0.86\columnwidth,trim={3.865cm 10.534cm 4.8260cm 10.793cm },clip]{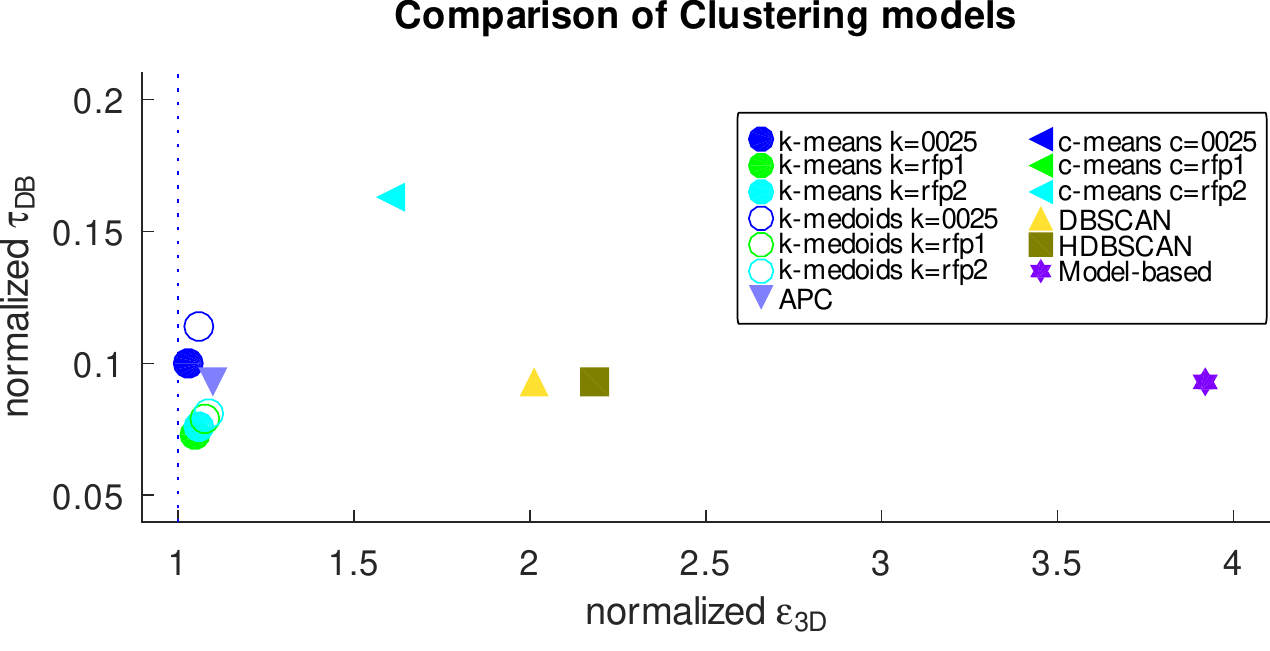}\\
    \includegraphics[width=0.86\columnwidth,trim={3.865cm 10.534cm 4.8260cm 11.163cm },clip]{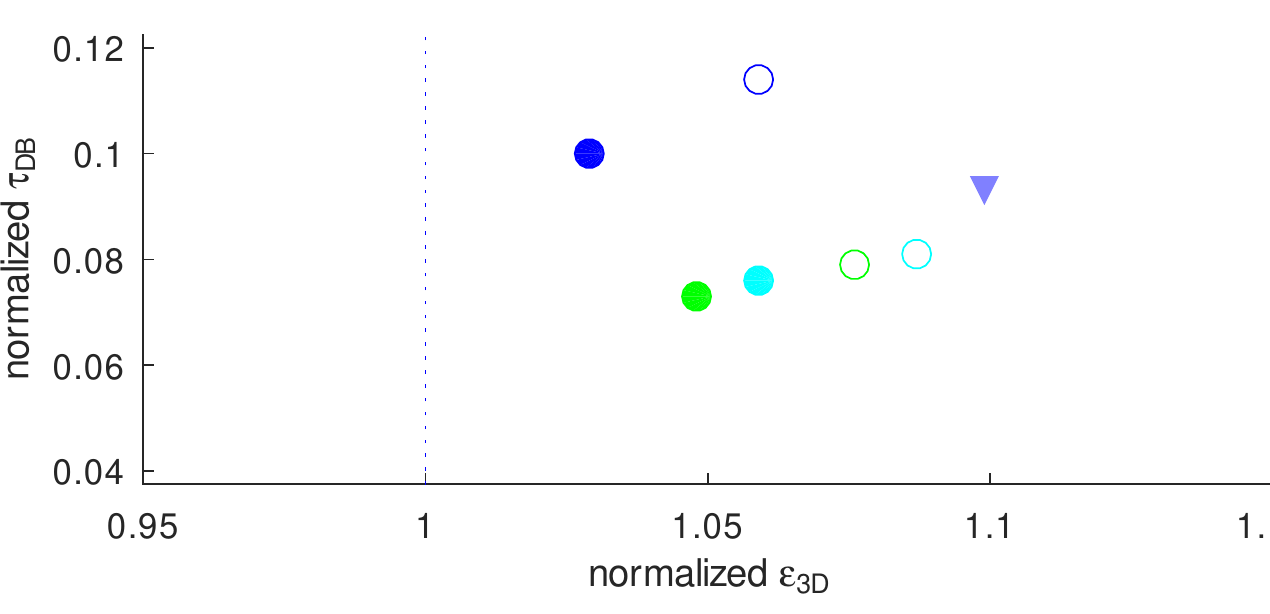}\\
    \caption{Visualization of the aggregated results reported by the clustering methods (top figure) and zoom to the best results (bottom figure)}
    \label{fig:clusteringresults}
\end{figure}

Table~\ref{tab:clusteringresults} and Fig.~\ref{fig:clusteringresults} show the results of some well-known clustering models ($k$-Means, $k$-Medoids, Fuzzy $c$-Means, Affinity Propagation, DBSCAN, HDBSCAN and Model-based)  in the literature. $k$-Means, $k$-Medoids and Fuzzy $c$-Means require the number of clusters as an input parameter. For them, we have tested three values: 25; the square root of the number of reference fingerprints in the radio map ($rfp1$); and the number of reference fingerprints in the radio map divided by 25 ($rfp2$). For DBSCAN-based methods, we used the optimal values for the minimum number of points and the distance used to locate the points in the neighbourhood. 

\begin{table}[!hbt]
    \centering
        \caption{Results reported by the selected clustering methods}
    \begin{tabular}{lcccccc}
    \toprule
    method&
    params&
    $\mywidetilde{\epsilon_{\text{3D}}} $ mean(std)&$\mywidetilde{\tau_{\text{DB}}}$ mean(std)\\
    \midrule
plain $1$-NN&--&	$1.00$ ($0.00$) &	$1.00$ ($0.00$) \\
$k$-means&$k=0025$&	$1.03$ ($0.03$) &	$0.10$ ($0.03$) \\
$k$-means&$k=rfp1$&	$1.05$ ($0.05$) &	$0.07$ ($0.04$) \\
$k$-means&$k=rfp2$&	$1.06$ ($0.06$) &	$0.08$ ($0.04$) \\
$k$-medoids&$k=0025$&	$1.06$ ($0.05$) &	$0.11$ ($0.04$) \\
$k$-medoids&$k=rfp1$&	$1.08$ ($0.07$) &	$0.08$ ($0.05$) \\
$k$-medoids&$k=rfp2$&	$1.09$ ($0.07$) &	$0.08$ ($0.04$) \\
$c$-means&$c=0025$&	$4.92$ ($8.92$) &	$0.20$ ($0.17$) \\
$c$-means&$c=rfp1$&	$4.16$ ($5.19$) &	$0.19$ ($0.13$) \\
$c$-means&$c=rfp2$&	$1.60$ ($0.84$) &	$0.16$ ($0.15$) \\
Affinity Propagation&--&	$1.10$ ($0.08$) &	$0.09$ ($0.04$) \\
DBSCAN&best params&	$2.01$ ($1.35$) &	$0.12$ ($0.11$) \\
HDBSCAN&best params&	$2.18$ ($3.37$) &	$0.31$ ($0.31$) \\
Model Based&--&	$3.92$ ($4.70$) &	$0.22$ ($0.30$) \\

    \bottomrule
    \end{tabular}
    \label{tab:clusteringresults}
\end{table}

\begin{figure}[!hbt]
    \centering
    \includegraphics[width=0.65\columnwidth]{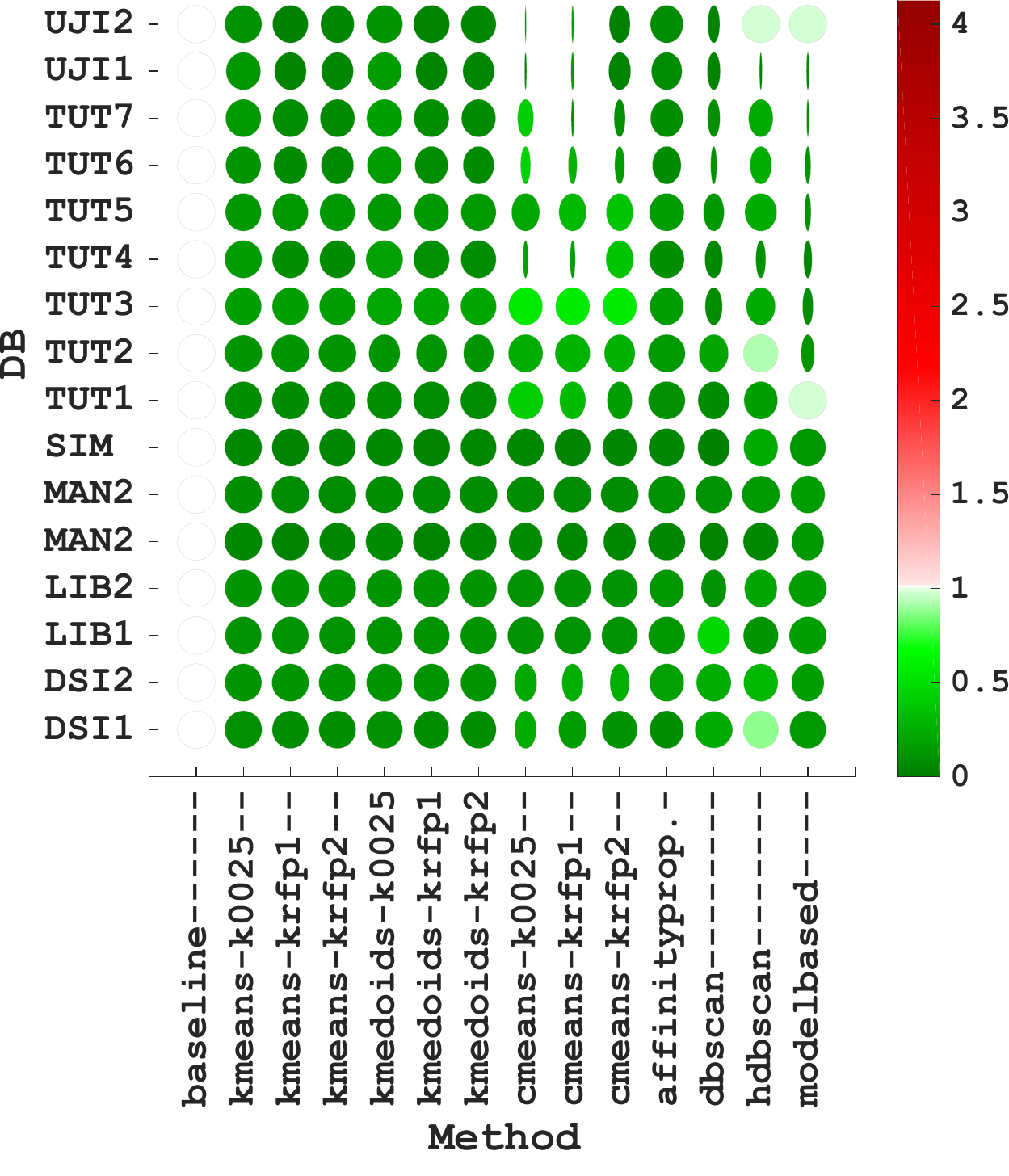}
    \caption{GMMS representation for the results provided by the clustering methods for the 16 datasets. The \emph{baseline} is the plain $1$-NN method.}
    \label{fig:clusteringdb}
\end{figure}

 The results reported in the table and figure show that all the clustering models reduce the computational time of fingerprinting. However, the aggregated computational time is high for HDBSCAN and Model Based as, in a few datasets, they failed at creating the clusters. Regarding the aggregated accuracy, we should discard DBSCAN, HDBSCAN, Model-based and $c$-Means as the aggregated positioning error is so high. The large variability might also indicate that those models do not work well in certain circumstances. In fact, according to the GMMS plot in Fig.~\ref{fig:clusteringdb}, they do not work for UJI~1 and some TUT datasets (vertical ellipsoids).
 
 In the bottom part of Fig.~\ref{fig:clusteringresults}, we can see that $k$-Means, $k$-Medoids and Affinity Propagation Clustering are very similar, being the $k$-means with $k = rfp1$ presenting the best trade-off between positioning error and execution time. The aggregated metrics have proven to be useful to compare method as a full analysis would be not possible. As an example, we provide the full results for $k$-means with $k = rfp1$ in Table \ref{table:fullresultskmeansrfp1}. 

\begin{table*}
\centering
\caption{Full results for $k$-Means with $k = rfp1$}
\label{table:fullresultskmeansrfp1}
\tabcolsep 2pt
\scriptsize
\begin{tabular}{ccccccccccccccccccccccccccc}
\toprule
&&\multicolumn{12}{c}{Mean Positioning Error -- $\epsilon^{i}_\text{3D}$ (m), $\mywidebar{\epsilon_\text{3D}}$ (m) and $\mywidehat{\epsilon_\text{3D}}$ (unitless) } 
&&\multicolumn{12}{c}{Execution Time -- $\tau^{i}_\text{DB}$ (s), $\mywidebar{\tau_\text{DB}}$ (s) and $\mywidehat{\tau_\text{DB}}$ (unitless) }  \\\cmidrule{3-14}\cmidrule{16-27}
Dataset&&
$\epsilon^{1}_\text{3D}$&
$\epsilon^{2}_\text{3D}$&
$\epsilon^{3}_\text{3D}$&
$\epsilon^{4}_\text{3D}$&
$\epsilon^{5}_\text{3D}$&
$\epsilon^{6}_\text{3D}$&
$\epsilon^{7}_\text{3D}$&
$\epsilon^{8}_\text{3D}$&
$\epsilon^{9}_\text{3D}$&
$\epsilon^{10}_\text{3D}$&
$\mywidebar{\epsilon_\text{3D}}$&
$\mywidehat{\epsilon_\text{3D}}$&&
$\tau^{1}_\text{DB}$&
$\tau^{2}_\text{DB}$&
$\tau^{3}_\text{DB}$&
$\tau^{4}_\text{DB}$&
$\tau^{5}_\text{DB}$&
$\tau^{6}_\text{DB}$&
$\tau^{7}_\text{DB}$&
$\tau^{8}_\text{DB}$&
$\tau^{9}_\text{DB}$&
$\tau^{10}_\text{DB}$&
$\mywidebar{\tau_\text{DB}}$&
$\mywidehat{\tau_\text{DB}}$\\
\midrule

{DSI1}&&    4.93&    4.97&    5.22&    5.21&    4.99&    5.40&    5.30&    5.29&    5.08&    4.85&    5.13&    1.04&  &    0.79&    0.86&    0.97&    0.86&    0.79&    0.81&    0.81&    1.01&    0.90&    0.91&    0.87&    0.07\\
{DSI2}&&    4.94&    5.13&    5.01&    5.25&    4.72&    4.88&    5.10&    5.75&    4.92&    4.78&    5.05&    1.02&  &    0.59&    0.52&    0.53&    0.54&    0.53&    0.53&    0.60&    0.52&    0.62&    0.58&    0.56&    0.11\\
{LIB1}&&    3.10&    3.16&    3.14&    3.10&    3.13&    3.11&    3.11&    3.12&    3.12&    3.19&    3.13&    1.04&  &    4.32&    4.77&    4.21&    4.17&    4.42&    4.17&    4.40&    4.50&    4.55&    4.33&    4.38&    0.09\\
{LIB2}&&    4.29&    4.18&    4.26&    4.54&    4.07&    4.27&    4.22&    4.19&    4.16&    4.42&    4.26&    1.02&  &    4.79&    5.64&    4.54&    4.47&    5.45&    4.62&    4.98&    4.92&    5.67&    4.81&    4.99&    0.11\\
{MAN1}&&    2.85&    2.89&    2.82&    2.88&    2.95&    2.84&    2.97&    2.84&    2.94&    2.82&    2.88&    1.02&  &    2.96&    3.08&    2.92&    2.92&    2.82&    2.89&    2.95&    3.02&    2.98&    2.92&    2.95&    0.02\\
{MAN2}&&    2.62&    2.46&    2.48&    2.56&    2.45&    2.43&    2.40&    2.60&    2.35&    2.45&    2.48&    1.01&  &    0.96&    0.93&    0.92&    1.04&    0.98&    0.94&    0.88&    0.92&    0.98&    1.02&    0.96&    0.07\\
{SIM}&&    3.27&    3.28&    3.36&    3.33&    3.28&    3.35&    3.31&    3.27&    3.37&    3.35&    3.32&    1.03&  &    5.00&    4.95&    4.93&    5.02&    4.88&    4.88&    4.98&    4.94&    4.89&    4.85&    4.93&    0.02\\
{TUT1}&&   10.06&    9.43&    9.76&   10.03&    8.99&   10.12&   10.77&    9.79&    9.37&   10.36&    9.87&    1.03&  &    1.25&    1.15&    1.11&    1.11&    1.06&    1.15&    1.39&    1.17&    1.12&    1.19&    1.17&    0.06\\
{TUT2}&&   13.84&   13.39&   16.02&   12.42&   13.98&   14.19&   13.33&   14.35&   13.91&   16.83&   14.22&    0.99&  &    0.29&    0.33&    0.29&    0.29&    0.31&    0.30&    0.28&    0.35&    0.30&    0.30&    0.30&    0.11\\
{TUT3}&&    9.96&   10.02&   10.02&   10.10&    9.92&    9.86&   10.14&    9.88&   10.05&    9.94&    9.99&    1.04&  &   16.99&   11.62&   13.30&   10.79&   13.69&   13.29&   12.20&   13.28&   11.52&   13.25&   12.99&    0.16\\
{TUT4}&&    6.62&    6.74&    6.64&    6.67&    6.74&    6.48&    6.54&    6.60&    6.59&    6.69&    6.63&    1.04&  &    5.12&    5.31&    5.12&    4.84&    8.17&    4.82&    5.50&    4.88&    6.08&    4.94&    5.48&    0.07\\
{TUT5}&&    7.74&    7.29&    7.14&    7.13&    7.12&    7.50&    7.51&    7.40&    7.37&    7.10&    7.33&    1.06&  &    1.36&    1.30&    1.49&    1.53&    1.39&    1.37&    1.46&    1.63&    1.35&    1.62&    1.45&    0.12\\
{TUT6}&&    2.25&    2.14&    2.20&    2.11&    2.17&    2.19&    2.21&    2.27&    2.20&    2.13&    2.19&    1.13&  &   37.47&   31.15&   35.34&   45.38&   40.16&   33.18&   40.99&   39.89&   33.41&   41.70&   37.87&    0.06\\
{TUT7}&&    2.91&    2.84&    2.92&    2.87&    2.92&    2.90&    2.88&    2.87&    2.84&    2.93&    2.89&    1.07&  &   30.48&   42.74&   29.63&   27.72&   33.01&   31.98&   31.57&   34.87&   29.53&   35.86&   32.74&    0.06\\
{UJI1}&&   12.76&   12.49&   13.01&   13.10&   12.28&   12.78&   12.72&   12.85&   13.06&   13.23&   12.83&    1.19&  &   11.75&   15.42&   12.34&   12.76&   12.40&   11.61&   13.61&   13.38&   10.52&   11.95&   12.57&    0.02\\
{UJI2}&&    8.72&    8.36&    8.89&    8.54&    8.43&    8.40&    8.51&    8.36&    8.53&    8.61&    8.54&    1.06&  &   43.39&   50.69&   48.61&   44.23&   44.27&   49.47&   44.85&   44.20&   49.26&   47.02&   46.60&    0.02\\
\bottomrule
\multicolumn{12}{r}{\multirow{2}{*}{$\mywidetilde{\epsilon_\text{3D}}$}}&mean&1.05&&
\multicolumn{10}{r}{\multirow{2}{*}{$\mywidetilde{\tau_\text{3D}}$}}&mean&0.07\\
\multicolumn{12}{r}{}&std&(0.05)&&
\multicolumn{10}{r}{}&std&(0.04)\\
\end{tabular}
\end{table*}

Despite the fact that a better analysis can be performed with the full results (see Table~\ref{table:fullresultskmeansrfp1}), one can focus on them after filtering out those methods that do not provide good general results. After selecting $k$-means with $k = rfp1$ as best choice, we can see that the model provides relative good results except for dataset UJI~1. In general, the computational time reduction is significant in all datasets, especially in those that have large radio maps (UJI~1, UJI~2, SIM, and MAN~1). The weakest point of $k$-Means, even providing the best general results, is that its accuracy and execution time depend on the partition done over the radio map. The variability in the error between runs is evident in Table~\ref{table:fullresultskmeansrfp1}, where the superscript in the evaluation metrics stand for the trial/run. $k$-means with $k = rfp1$ has similar general accuracy that the baseline in the best run, whereas it is similar to affinity propagation (which provides deterministic clustering) in the worst.

\subsection{Data Compression}


The third use case, in which we show the benefits of the aggregated metrics, is based on studying the impact of the database compression using \ac{akm} algorithm~\cite{klus2020rss} on its positioning capabilities. In this use case, we would like to apply the aggregated metrics to first determine the optimal parametrisation of \ac{akm} and then to compare to the plain $1$-NN.

For the first part, to determine the optimal parameters, we cannot use the same baseline method for results normalization, as some metrics are dependent on the clustering approach selected. In particular,
the studied metrics are \ac{mse} after Stage 1 and Stage 2 of the \ac{akm} compression ($MSE_{S1}$ and $MSE_{S2}$) , Normalized Mean Positioning Error ($\mywidetilde{\epsilon_{\text{3D}}}$), and achieved \ac{cr}.

The metrics $MSE_{S1}$ and $MSE_{S2}$ correspond to the mean squared difference between the original RSSI values and the RSSI values after compression using clustering. We analyze the \ac{akm} algorithm by modifying the number of clusters $K$. The algorithm is initiated by clustering all train dataset samples’ \ac{rss} values separately, thus first creating a one-dimensional array of all entries of the train \ac{rss} samples. The algorithm further clusters the array into $K$ clusters, each of them defined by its centroid coordinate (a single number). The $MSE_{S1}$ is then calculated as \ac{mse} between the original test \ac{rss} data and its reconstruction after clustering using the previously obtained centroids. The algorithm then adapts the centroid coordinates by including the test \ac{rss} data, which results in shifting the centroid coordinates as described in~\cite{klus2020rss}. $MSE_{S2}$ is then calculated as \ac{mse} between the original test \ac{rss} data and its reconstruction using the adapted centroid coordinates. The algorithm then applies the $k$-\ac{nn} regression with $k$ = 1 on the clustered train and test datasets (where the original \ac{rss} values were replaced by the corresponding centroid coordinates) and obtains the positioning predictions for the test samples. 

The \ac{cr} is then calculated as number of bits of the original data divided by the number of bits of the compressed data. We assume that the uncompressed integer-valued \ac{rss} measurements are saved in 7-bit format (allowing 128 unique \ac{rss} values) and that $ceil(log_2(K))$ unique values can be saved with $K$ bits, where ceil() rounds up to the next higher integer. 

Therefore, for comparing the different setups of \ac{akm}, we have chosen the simplest version with $K = 2$ (for $k$ from $k$-Means) as baseline and, then, we normalize the four metrics to it.  
%
The aggregated results on all 16 considered datasets are shown in Table~\ref{tab:CompressionResults} and are achieved by averaging over 10 repetitions of the algorithm for each dataset per each $K$ setting. The considered $K$ values for \ac{akm} clustering are 2, 4, 7, 15, 25 and 35. As the number of clusters increases, the aggregated \ac{cr} decreases accordingly, depending on the number of required bits to compress each value. The aggregated $MSE_{S1}$ and $MSE_{S2}$ parameters also decrease with increasing $K$ parameter. This is a natural result of decreasing rounding error during clustering (larger number of clusters leads to smaller distance each sample is shifted during clustering). Finally, the aggregated positioning error $\mywidetilde{\epsilon_{\text{3D}}}$ also decreases.

\begin{table}[!hbt]
\caption{Results for setting the best overall parameters for \ac{akm}}
\label{tab:CompressionResults}
\begin{center}
\begin{tabular}{lccccc}
\toprule
$K$ & 
$\mywidetilde{MSE_{S1}}$ & 
$\mywidetilde{MSE_{S2}}$ & 
$\mywidetilde{\epsilon_{\text{3D}}}$ & 
$\mywidetilde{CR}$&
$\mywidetilde{\mathcal{F}}$\\ 
\midrule
2         & 1.000 (0.00) & 1.000 (0.00) & 1.00 (0.00)  & 1.00 (0.00) & 1.00 \\
4         & 0.163 (0.04) & 0.164 (0.04) & 0.84 (0.10)  & 0.50 (0.00) & 0.76\\
7         & 0.050 (0.02) & 0.051 (0.02) & 0.81 (0.11)  & 0.33 (0.00) & 0.73\\
15        & 0.010 (0.00) & 0.010 (0.00) & 0.79 (0.12)  & 0.25 (0.00) & \textbf{0.72} \\
25        & 0.003 (0.00) & 0.003 (0.00) & 0.79 (0.12)  & 0.20 (0.00) & 0.72\\
35        & 0.001 (0.00) & 0.001 (0.00) & 0.79 (0.12)  & 0.17 (0.00) & 0.73\\
\bottomrule
\end{tabular}
\end{center}
\end{table}

However, given the results reported in the table, it is not that easy to retrieve a winning setup, as larger $K$ values (for $k$-Means) lead to better results but lower compression.
After some discussion, we decided for that particular problem an additional aggregation in the way:
\begin{equation}
\mywidetilde{\mathcal{F}} =
    0.05\cdot\mywidetilde{MSE_{S1}} +
    0.05\cdot\mywidetilde{MSE_{S2}} +\ 
    0.9\cdot\left(\mywidetilde{\epsilon_{\text{3D}}}\right)^2 + 
    0.2\cdot(1-\mywidetilde{CR})\nonumber
\end{equation}
where the aggregated positioning error has more weight than the \acf{cr} and the two metrics based on the \ac{mse}, becoming $K=15$ the best configuration for \ac{akm}.

Table~\ref{tab:CompressionResults2} compares the plain $1$-NN with \acf{akm} and $k=15$, using $1$-NN as baseline. The results show that the selection of parameters led \ac{akm} to provide similar accuracy as $1$-NN with a more efficient RSSI representation. 

\begin{table}[!hbt]
\caption{Comparison of plain $1$-NN with \ac{akm}}
\label{tab:CompressionResults2}
\begin{center}
\begin{tabular}{lcccc}
\toprule
method & 
$\mywidetilde{\epsilon_{\text{3D}}}$ mean(std)& 
$\mywidetilde{CR}$ mean(std)\\ 
\midrule
Plain $1$-NN.     & 1.00 (0.00) & 1.00 (0.00)\\
\ac{akm} ($k=15$) & 1.00 (0.03) & 1.75 (0.00)\\
\bottomrule
\end{tabular}
\end{center}
\end{table}

\begin{table}[hbtp]
\centering
\caption{Full results of \ac{akm}. }
\label{table:fullakm}
\aboverulesep = 0.3025mm
\belowrulesep = 0.482mm
\begin{tabular}{lSSccc}   
\toprule
&\multicolumn{2}{c}{Absolute values}&
&\multicolumn{2}{c}{Norm. values}\\\cmidrule{2-3}\cmidrule{5-6}
Dataset& 
{$\mywidebar{\epsilon_{\text{3D}}}$ (m)} & {$\mywidebar{CR}$ (s)} && {$\mywidehat{\epsilon_{\text{3D}}}$}& {$\mywidehat{CR}$}  
\\\midrule
DSI~1&4.88&1.75&&0.99&1.75\\
DSI~2&5.21&1.75&&1.02&1.75\\
LIB~1&3.04&1.75&&1.01&1.75\\
LIB~2&4.22&1.75&&1.01&1.75\\
MAN~1&2.88&1.75&&1.02&1.75\\
MAN~2&2.39&1.75&&0.97&1.75\\
SIM&3.60&1.75&&1.10&1.75\\
TUT~1&9.75&1.75&&1.02&1.75\\
TUT~2&14.25&1.75&&0.99&1.75\\
TUT~3&9.55&1.75&&1.00&1.75\\
TUT~4&6.35&1.75&&1.00&1.75\\
TUT~5&6.98&1.75&&1.01&1.75\\
TUT~6&1.98&1.75&&1.02&1.75\\
TUT~7&2.71&1.75&&1.01&1.75\\
UJI~1&10.21&1.75&&0.94&1.75\\
UJI~2&7.92&1.75&&0.98&1.75\\
\bottomrule
\end{tabular}
\end{table}

\section{Discussion \& Conclusions}
\label{sec:conclusions}
In this paper, the importance of evaluating \acp{ips} in multiple scenarios was discussed. This is an essential step for the characterization of the true performance of an \ac{ips}, and by consequence, the fair comparison with other solutions. However, due to the complexity involved in preparing experiments in multiple scenarios, dataset publishing by the research community is of utmost importance. Moreover, considering different performance metrics over multiple scenarios, an aggregation of the evaluation metrics is proposed to enable high-level comparison of \acp{ips}.  

When dealing with an evaluation that involves multiple metrics, an interesting approach to explore is their combination into a single metric. In such a case, we suggest to apply the following weighted combination of the aggregated metrics:
 \begin{equation}
     \tilde{\mathcal{F}}_{method} = 
     \omega_{{\mathcal{M}}}\cdot\tilde{\mathcal{M}}_{method}+
     \ldots+
     \omega_{{\mathcal{Q}}}\cdot\tilde{\mathcal{Q}}_{method}
 \end{equation}
where the weight values ($\omega$) can be user-defined. 
 A weighted combination allows taking into consideration the requirements of a particular use case or \ac{ips} deployment, being possible to define how important each of the performance metrics is for the overall evaluation. An \ac{ips} may be the best for one use case but not for another with different requirements.  
 
The proposed aggregation of evaluation metrics simplifies complex comparisons between \acp{ips} when considering many parameters. In addition, it also simplifies the process of selecting the best \ac{ips} for a specific scenario or deployment.
 
 The use cases we described in this work give us the possibility of validating our proposal to show how multiple datasets can improve the \acp{ips} evaluation. This work highlights also the need of a common and shared graphical representation able to give immediately the impression of what kind of methods are better in all the considered datasets. In addition, we were able to provide results on three different independent experiments in an 8-page paper, demonstrating also the power of the aggregated metrics to summarize general results in a compact table or figure. 
 
 However, the evaluation of an \ac{ips} is not easy when multiple metrics need to be considered.  The combination of several metrics such as accuracy, installation complexity, user acceptance, availability and integrability as done in \cite{6401100} would need further discussion. Nevertheless, we suggest a new level of aggregation based on a user-defined weighted combination. 
 
 Finally, we have found two major issues in the literature. The first one is the lack of guidelines to prepare, collect and publish datasets for indoor positioning. That could be a reason why the community is not adopting datasets in their evaluation as they may not be interoperable with their research. The second one is that most datasets are RSSI-based, the community needs datasets covering other positioning technologies. As the proposed aggregation is agnostic to the positioning technology.




\section*{Acknowledgment}

We would like to thank Germán Martín Mendoza-Silva and Philipp Richter for their invaluable advice, which encouraged us to prepare the current work presented in this paper. 

\section*{CRediT Statement}

\noindent \textbf{J. Torres-Sospedra:}
Conceptualization Ideas; 
Methodology	Development;
SW Programming, 
Validation;
Formal analysis;
Investigation;
Resources;
Data Curation;
Writing (Original Draft, Review \& Editing); and
Supervision;

\noindent \textbf{I. Silva:}
Discussion and Writing (Review \& Editing).

\noindent \textbf{L. Klus:}
Methodology	Development;
Software Programming, 
Validation;
Investigation;
Writing (Review \& Editing);

\noindent \textbf{D. Quezada-Gaibor:}
Software Programming, 
Validation;
Investigation;
Writing (Review \& Editing);

\noindent \textbf{A. Crivello:}
Methodology	Development;
Formal analysis;
Investigation and
Writing (Original Draft, Review \& Editing).

\noindent \textbf{P. Barsocchi:}
Conceptualization Ideas; 
Investigation;
Writing (Original Draft, Review \& Editing); 
Supervision.

\noindent \textbf{C. Pendão:}
Conceptualization Ideas and Writing (Original Draft, Review \& Editing);

\noindent \textbf{E. S. Lohan:}
Methodology Development;
Resources;
Writing (Review \& Editing); and
Supervision;

\noindent \textbf{J. Nurmi:}
Conceptualization Ideas;
Resources;
Writing (Review \& Editing).

\noindent \textbf{A. Moreira:}
Conceptualization Ideas,
Methodology	Development and Supervision

\balance
\renewcommand*{\UrlFont}{\rmfamily}
\printbibliography

\end{document}